\documentclass[conference,a4paper]{IEEEtran}
\usepackage{xcolor}
\usepackage{balance}
\usepackage{cite}
\usepackage{multirow}
\usepackage[pdftex]{graphicx}
\usepackage{amsmath,amssymb,amsfonts}
\usepackage{textcomp}
\usepackage[caption=false,font=footnotesize]{subfig}
\usepackage[nolist]{acronym}
\usepackage[capitalize]{cleveref}
\usepackage{nicefrac}
\usepackage{bm}
\usepackage{graphicx} 
\crefname{table}{Tab.}{Tabs.}
\Crefname{table}{Tab.}{Tabs.}
\crefformat{equation}{(#2#1#3)}

\newcommand{\TRx}[3]{#1_{\rm #2}^{\left( {\rm #3} \right)} }

\begin{document}
\begin{acronym}

\acro{5G}[5G]{fifth generation}
\acro{6G}[6G]{sixth generation}
\acro{3GPP}[3GPP]{3rd Generation Partnership Project}

\acro{AB}[AB]{alpha-beta}
\acro{ABG}[ABG]{alpha-beta-gamma}
\acro{ADC}[ADC]{analog-to-digital converter}
\acro{AoD}[AoD]{angle-of-departure}
\acro{AoA}[AoA]{angle-of-arrival}
\acro{AWG}[AWG]{Arbitrary Waveform Generator}
\acro{AWGN}[AWGN]{additive white Gaussian noise}

\acro{BER}[BER]{bit error ratio}

\acro{CDF}[CDF]{cumulative distribution function}
\acro{CI}[CI]{close-in}
\acro{CIR}{channel impulse response}
\acro{CW}[CW]{continuous wave}
\acro{CSIT}{channel state information at the transmitter}
\acro{CSI}{channel state information}
\acro{CTF}{channel transfer function}
\acro{CTF}[CTF]{channel transfer function}

\acro{D2D}[D2D]{device-to-device}
\acro{DAC}[DAC]{digital-to-analog converter}
\acro{DFT}[DFT]{discrete Fourier transform}
\acro{DPSS}[DPSS]{discrete prolate spheroidal sequences}
\acro{DR}[DR]{dynamic range}
\acro{DSD}{Doppler power spectral density}

\acro{EVD}[EVD]{eigenvalue decomposition}
\acro{eMBB}[eMBB]{enhanced mobile broadband}
\acro{EM}[EM]{expectation–maximization}

\acro{FDD}{frequency-division duplex}
\acro{FPGA}[FPGA]{Field Programmable Gate Array}
\acro{FR1}[FR1]{frequency range 1}
\acro{FR2}[FR2]{frequency range 2}
\acro{FR3}[FR3]{frequency range 3}
\acro{FSPL}[FSPL]{free space path loss}

\acro{GRU}{Gated Recurrent Unit}

\acro{HPBW}[HPBW]{half-power beamwidth}
\acro{HRPE}[HRPE]{high-resolution parameter estimation}
\acro{HST}[HST]{High-Speed Train}

\acro{IC}[IC]{Integrated Circuit}
\acro{ICI}[ICI]{Inter-Carrier Interference}
\acro{IDFT}[IDFT]{inverse discrete Fourier transform}
\acro{IF}[IF]{intermediate frequency}
\acro{IFFT}{Inverse Fast Fourier Transform}
\acro{ISI}[ISI]{Inter-Symbol Interference}
\acro{ITS}[ITS]{intelligent transportation systems}
\acro{ISI}[ISI]{Inter-Symbol Interference}

\acro{LMMSE}[LMMSE]{linear minimum mean square error}
\acro{LNA}[LNA]{low-noise amplifier}
\acro{LO}[LO]{local oscillator}
\acro{LOS}[LOS]{line-of-sight}
\acro{LSF}[LSF]{local scattering function}
\acro{LSTM}{long short-term memory}
\acro{LS}[LS]{least-squares}

\acro{MAC}[MAC]{medium access control}
\acro{MIMO}[MIMO]{multiple-input multiple-output}
\acro{ML}{machine learning}
\acro{MMW}[mmWave]{millimeter wave}
\acro{MoM}[MoM]{method-of-moments}
\acro{MPC}[MPC]{multipath component}
\acro{MRC}[MRC]{maximal ratio combining}
\acro{MSE}[MSE]{mean squared error}
\acro{MUSIC}[MUSIC]{multiple signal classification}

\acro{NR}[NR]{new radio}
\acro{NLOS}[NLOS]{non-line-of-sight}

\acro{OFDM}[OFDM]{orthogonal frequency-division multiplexing}
\acro{OOBA-BD}[OOBA-BD]{out-of-band aided beam-domain}
\acro{OBABE}[OBABE]{out-of-band aided beam-domain estimation}
\acro{OLOS}[OLOS]{obstructed line-of-sight}

\acro{PA}[PA]{power amplifier}
\acro{PAS}{power angular spectrum}
\acro{PDP}{power delay profile}
\acro{PCB}[PCB]{Printed Circuit Board}
\acro{PLE}[PLE]{path-loss exponent}

\acro{RF}[RF]{radio frequency}
\acro{RMS}[RMS]{root-mean-square}
\acro{RMSE}[RMSE]{root mean squared error}
\acro{RNN}{recurrent neural networks}
\acro{Rx}[Rx]{receiver}

\acro{QAM}[QAM]{quadrature amplitude modulation}

\acro{SAGE}[SAGE]{space-alternating generalized expectation maximization}
\acro{SE}[SE]{spectral efficiency}
\acro{SISO}[SISO]{single-input single-output}
\acro{SIMO}[SIMO]{single-input multiple-output}
\acro{SINR}[SINR]{signal-to-interference-and-noise ratio}
\acro{SMD}[SMD]{Surface Mount Device}
\acro{SNR}[SNR]{signal-to-noise ratio}
\acro{SVD}[SVD]{singular value decomposition}
\acro{SVM}[SVM]{support vector machine}

\acro{TDD}{time-division duplex}
\acro{TDL-A}[TDL-A]{tapped delay line A}
\acro{TDL-D}[TDL-D]{tapped delay line D}
\acro{Tx}[Tx]{transmitter}

\acro{URLLC}[URLLC]{ultra-reliable low-latency communication}
\acro{ULA}[ULA]{uniform linear array}
\acro{UPA}[UPA]{uniform planar array}

\acro{V2I}[V2I]{vehicle-to-infrastructure}
\acro{V2V}[V2V]{vehicle-to-vehicle}
\acro{V2X}[V2X]{vehicle-to-everything}
\acro{VNA}[VNA]{vector network analyzer}

\end{acronym}

\title{Multi-Band Indoor MIMO Channel Measurements Across Sub-6\,GHz, Mid-Band and mmWave: \\ Comparing 2.55\,GHz, 12.75\,GHz and 25.5\,GHz}

\author{\IEEEauthorblockN{
Faruk Pasic\IEEEauthorrefmark{1},
Florian Kiss\IEEEauthorrefmark{1},
Florian Asperger\IEEEauthorrefmark{1}\IEEEauthorrefmark{2},
Diego Dupleich\IEEEauthorrefmark{3},
Stefan Schwarz\IEEEauthorrefmark{1} and
Markus Rupp\IEEEauthorrefmark{1}
}%

\IEEEauthorblockA{\IEEEauthorrefmark{1}
Institute of Telecommunications, TU Wien, Vienna, Austria}
\IEEEauthorblockA{\IEEEauthorrefmark{2}
Christian Doppler Laboratory for Digital Twin assisted AI for Sustainable Radio Access Networks, Vienna, Austria}
\IEEEauthorblockA{\IEEEauthorrefmark{3}
Institute for Information Technology, TU Ilmenau, Ilmenau, Germany} 
\IEEEauthorblockA{faruk.pasic@tuwien.ac.at}
}

\IEEEoverridecommandlockouts 

\makeatletter
\def\thanks#1{\protected@xdef\@thanks{\@thanks
        \protect\footnotetext{#1}}}
\makeatother

\maketitle

\begin{abstract}
Future wireless systems for indoor applications are expected to extend their operating frequencies from sub-6\,GHz to mid-band and \ac{MMW} frequencies to support higher data rates. 
To characterize the frequency-dependent propagation behavior across these bands, multi-band channel measurements are essential. 
In this work, we conduct double-directional \ac{MIMO} channel measurements in an indoor environment to enable a consistent comparison across three frequency bands. 
Measurements are performed using a 16$\times$16 \ac{MIMO} configuration at center frequencies of 2.55\,GHz, 12.75\,GHz and 25.5\,GHz, covering both \ac{LOS} and \ac{OLOS} conditions. 
The measured channels are processed using the \ac{SAGE} algorithm to extract multipath parameters, which are subsequently used to characterize the angular power distributions across the three frequency bands. 
The results show substantial angular-domain similarity across all three bands in both \ac{LOS} and \ac{OLOS} conditions, while also highlighting frequency-dependent variations.
\end{abstract}
\vskip0.5\baselineskip

\acresetall

\section{Introduction} \label{sec:introduction}
Global mobile data traffic has reached approximately 150\,exabytes per month and is expected to continue increasing rapidly, reaching around 500\,exabytes per month by 2030~\cite{ericsson2024}.
A significant share of this growth is driven by emerging indoor wireless applications, such as ultra-high-definition video streaming, high-resolution virtual and augmented reality (VR/AR) and environmental monitoring~\cite{Kanhere2025}.
One approach to meet this growing demand for higher data rates is to exploit additional spectrum for wireless communications.
In recent years, the mid-band spectrum, spanning approximately 6\,GHz -- 24\,GHz and referred to as \ac{FR3}~\cite{3GPP_TR_38_820}, has attracted increasing interest for wireless communications~\cite{Shakya2024}. 
\ac{FR3} complements the established sub-6\,GHz \ac{FR1} and \ac{MMW} \ac{FR2}, which covers 24\,GHz -- 300\,GHz~\cite{Rappaport2013, Molisch2025, Hammoud2025_TVT}.
The growing interest in \ac{FR3} therefore motivates detailed investigations of propagation characteristics in this frequency range, particularly in indoor environments.

To understand the evolution of indoor propagation characteristics across frequency, measurements in the mid-band alone are not sufficient, and measurements at sub-6\,GHz and \ac{MMW} frequencies are also required for a consistent comparison.
Several measurement campaigns have investigated mid-band propagation exclusively~\cite{Shakya2024, Richard2025, Serghiou2025, Das2026}.
In addition, numerous multi-band indoor measurement-based studies have considered sub-6\,GHz and mmWave frequencies~\cite{Anderson2004, Koymen2015, DeJong2018, Jiang2020, Pasic2023, Pasic2025_OJCOMS, Radovic2023}, while some have investigated mid-band together with \ac{MMW}~\cite{Dupleich2019_indoor}.
However, only a limited number of indoor measurement campaigns have considered all three frequency ranges~\cite{Virk2017, Zhang_G_2021, Huang2016}, highlighting the need for further comprehensive measurement studies.
Since channel characteristics fundamentally determine the performance limits of wireless communication systems, understanding their evolution across frequency is essential for the development of accurate channel models.

A meaningful comparison across frequency bands requires carefully designed measurements under consistent propagation conditions. 
In particular, the relative positioning of antennas or antenna arrays across frequency bands is an important but often overlooked aspect. 
Even a physical separation of a few centimeters can correspond to several wavelengths at \ac{MMW} frequencies, introducing spatial offsets that may alter small-scale fading and bias the resulting channel statistics. 

\textbf{Contribution:}
In this paper, we conduct double-directional 16$\times$16 \ac{MIMO} channel measurements in an indoor environment at 2.55\,GHz, 12.75\,GHz and 25.5\,GHz, covering the three considered frequency bands.
Measurements are performed under both \ac{LOS} and \ac{OLOS} conditions using a consistent measurement geometry. 
The extracted multipath channel parameters are then analyzed to compare the angular power distributions across the three frequency bands.

\textbf{Organization:}
The remainder of the paper is organized as follows. 
We describe our multi-band measurement setup in~\cref{sec:measurement_setup}.
The measurement campaign and the considered scenarios are presented in~\cref{sec:measurement_campaign}.
We provide and discuss the results of the measured wireless channel in~\cref{sec:results} and conclude our work in~\cref{sec:conclusion}.

\begin{figure*}[t]
	\centering
	\subfloat[\textbf{Sub-6\,GHz/Mid-band:} The signal is generated and analyzed directly at a center frequency of either $\TRx{f}{c}{l}=2.55$\,GHz or $\TRx{f}{c}{m}=12.75$\,GHz, and transmitted to and received from the antennas via cables.]{
		\includegraphics[width=2\columnwidth]{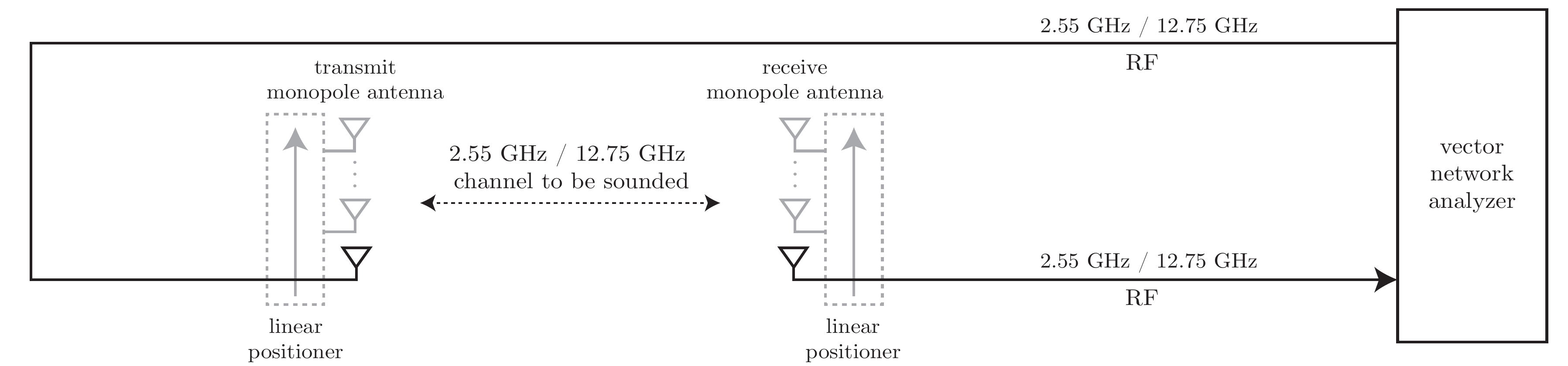}
		\label{fig:hardware_setup_lower_mid_band}
	}
    \\
	\subfloat[\textbf{Millimeter-wave:} The signal is generated and analyzed at an intermediate frequency (IF) of $\TRx{f}{IF}{h}=3.5$\,GHz, up-converted to a center frequency of $\TRx{f}{c}{h}=25.5$\,GHz for transmission and subsequently down-converted at the receiver.]{
		\includegraphics[width=2\columnwidth]{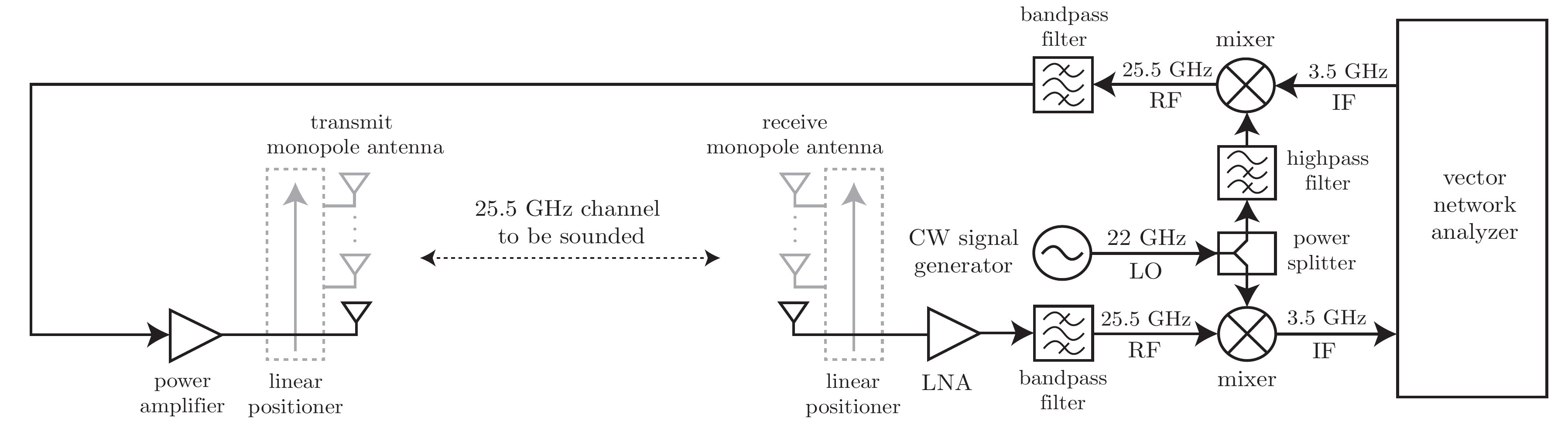}
		\label{fig:hardware_setup_higher_band}
	}
	\caption{Our multi-band measurement setup utilizes linear positioners that move along a rectangular grid in the horizontal plane at both the transmitter and receiver to realize a virtual multi-antenna configuration.}
	\label{fig:block_diagram}
\end{figure*}

\section{Measurement Setup} \label{sec:measurement_setup}
Our measurement setup employs 16$\times$16 virtual \ac{MIMO} configuration and covers three different frequency bands: sub-6\,GHz, mid-band and \ac{MMW}.
At the core of the measurement setup is  a \ac{VNA} (R\&S ZVA24), which is used to acquire the complex-valued \ac{CTF} of the propagation channel.
At both the transmitter and receiver, the virtual \ac{MIMO} antenna array is realized using a single antenna mounted on a 1.5-meter-high support pole. 
This support pole is attached to a stepper-motor-controlled linear positioner, which moves the antenna along a rectangular grid to synthesize a virtual 4$\times$4 \ac{UPA} in the horizontal plane.
The virtual \acp{UPA} at the transmitter and receiver share the same geometric center across all frequency bands, enabling a fair comparison of the measured channels~\cite{Pasic2026_iWAT}.
For each frequency band, an omnidirectional monopole antenna is used at both the transmitter and receiver.
The separation between adjacent virtual antenna elements is set to 0.4\,$\TRx{\lambda}{}{b}$, where $\TRx{\lambda}{}{b}$ denotes the wavelength of the corresponding center frequency. 
Throughout this paper, the superscript $\left( \cdot \right)^{\left( \rm b \right)}$ denotes a frequency-band-dependent quantity, where ${\rm b} \in \{ {\rm l}, {\rm m}, {\rm h} \}$.
Here, ${\rm l}$ denotes the lower (sub-6\,GHz) band, ${\rm m}$ denotes the mid-band and ${\rm h}$ denotes the higher (\ac{MMW}) band.
To ensure a fair comparison across the three bands, monopole antennas with comparable radiation patterns are employed.
A detailed description of the setup configuration for each frequency band is provided below.

\textbf{Sub-6\,GHz/Mid-band:}
For the sub-6\,GHz band and mid-band measurements, the signals are generated and captured directly by the \ac{VNA} at center frequencies of $\TRx{f}{c}{l}=2.55$\,GHz and $\TRx{f}{c}{m}=12.75$\,GHz, as illustrated in~\cref{fig:hardware_setup_lower_mid_band}.
The transmit signal generated by the \ac{VNA} is fed through a coaxial cable to the transmit monopole antenna.
At the receiver, the signal captured by the receive monopole antenna is routed through a coaxial cable to the \ac{VNA} for acquisition.

\begin{figure*}[t]
	\centering
	\includegraphics[width=1.97\columnwidth]{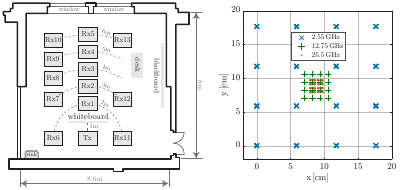}
	\caption{Measured indoor environment (left) and Tx/Rx virtual UPA configurations with identical geometric centers across all three frequency bands (right).}
	\label{fig:environment}
\end{figure*}

\begin{figure}[t]
	\centering
    \includegraphics[width=\columnwidth]{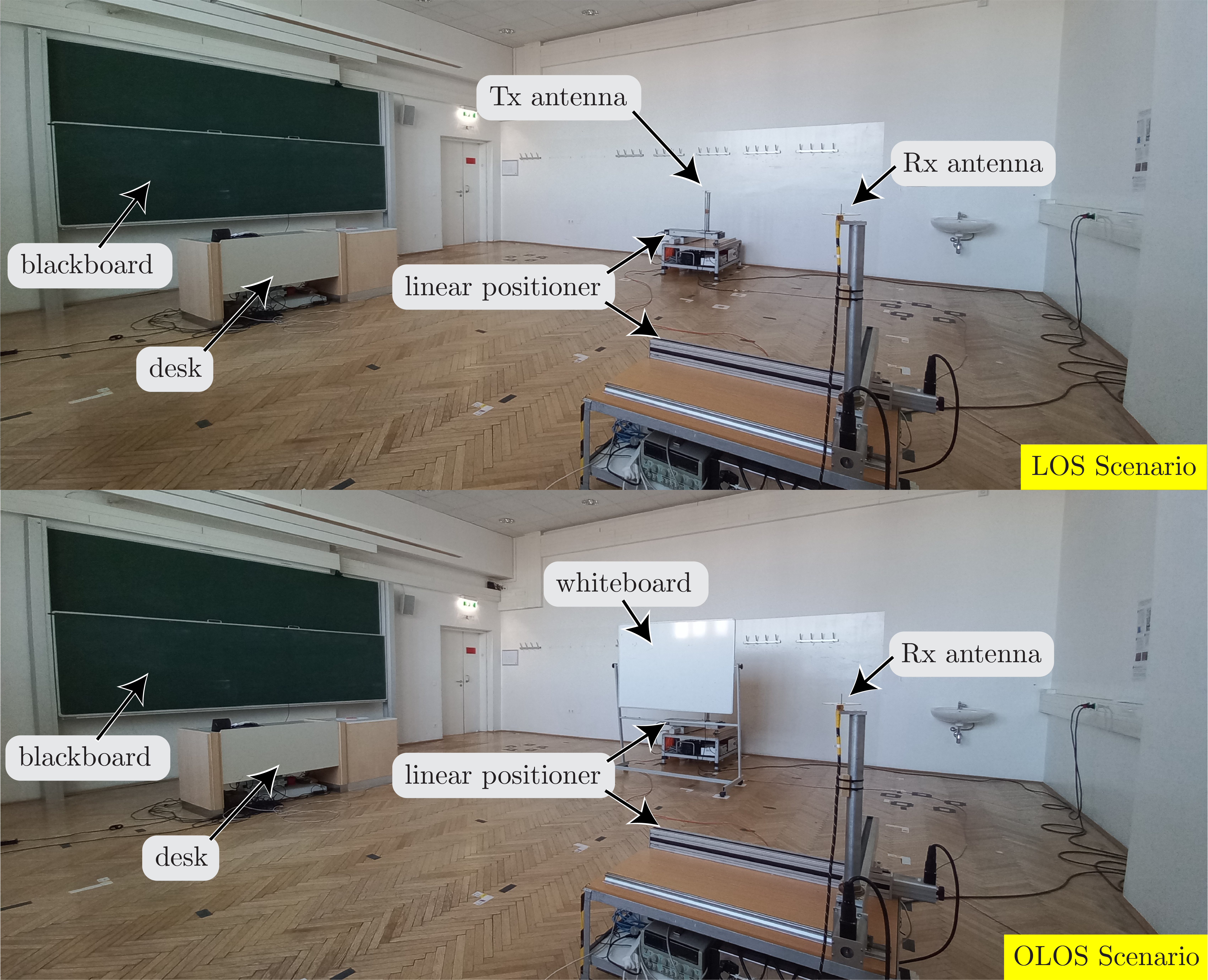}
	\caption{Lecture room under LOS conditions (top) and OLOS conditions with a whiteboard placed between the Tx and Rx (bottom).}
	\label{fig:LOS_OLOS_scenarios}
\end{figure}

\textbf{Millimeter-Wave:}
For the \ac{MMW} measurements, direct signal transmission at the desired center frequency is not feasible due to the frequency limitation of the employed \ac{VNA}, which support signal frequencies only up to 24\,GHz.
Therefore, frequency conversion is employed at both the transmitter and receiver, as illustrated in~\cref{fig:hardware_setup_higher_band}.

At the transmitter, the \ac{IF} signal generated by the \ac{VNA} is up-converted using a mixer, with the \ac{LO} signal is provided by a \ac{CW} signal generator (R\&S SMF100A).
Specifically, an \ac{IF} of $\TRx{f}{IF}{h}=3.5$\,GHz and an \ac{LO} frequency of $\TRx{f}{LO}{h}=22$\,GHz result in an \ac{RF} center frequency of $\TRx{f}{c}{h}=25.5$\,GHz.
The up-converter is implemented using a mixer (Mini-Circuits ZMDB-44H-K+), followed by a bandpass filter (Techniwave TWBF201221A) to suppress unwanted spectral components.
The filtered signal is subsequently amplified using a \ac{PA} (RF-Lambda RFLUPA18G47GC) to compensate for the increased propagation and hardware losses at \ac{MMW} frequencies.
The amplified signal is then fed through a coaxial cable to the transmit monopole antenna.

At the receiver, the signal captured by the receive monopole antenna is first amplified using an \ac{LNA} (RF-Lambda R24G40GSB) and subsequently filtered using the same type of bandpass filter (Techniwave TWBF201221A). 
The filtered signal is then down-converted using a mixer (Mini-Circuits ZMDB-44H-K+), with the \ac{LO} signal again provided by the R\&S SMF100A.
The resulting \ac{IF} signal is finally fed to the \ac{VNA} for acquisition.
Note that the same \ac{CW} signal generator provides the \ac{LO} signal for both the up- and down-conversion stages. 
The generated \ac{LO} signal is divided using a power splitter (Mini-Circuits ZC4PD-18263-S+) to provide the two conversion paths.
On the down-conversion path, a high-pass filter (Mini-Circuits ZHSS-11G-S+) is used to suppress undesired coupling of the \ac{IF} transmit and receive signals through the \ac{LO} path.
Finally, for the \ac{MMW} case, a back-to-back calibration is performed by directly connecting the transmitter and receiver via attenuators to account for and remove the effects of the \ac{RF} components in the signal path.

\section{Measurement Campaign} \label{sec:measurement_campaign}

\begin{figure*}[t]
	\centering
	\subfloat[LOS case: The direct path between the Tx and Rx is the strongest component across all three frequency bands.]{
		\includegraphics[width=2\columnwidth]{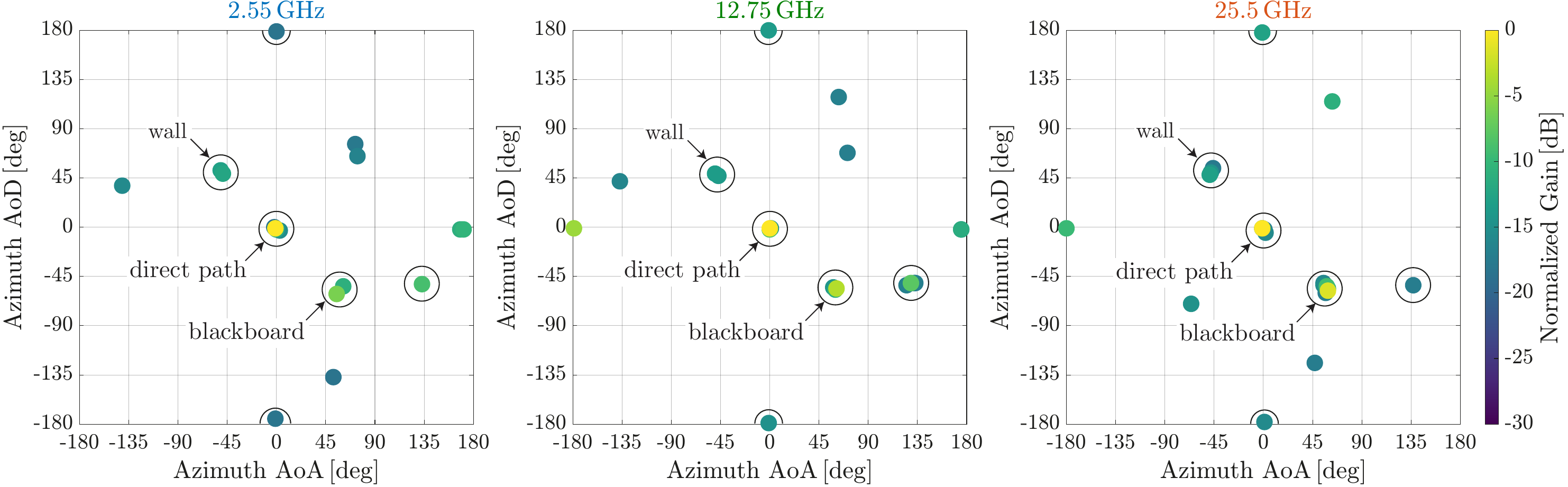}
		\label{fig:SAGE_AoA_vs_AoD_LOS}
	}
    \\
	\subfloat[OLOS case: The reflection from the blackboard is the strongest resolved component across all three frequency bands.]{
		\includegraphics[width=2\columnwidth]{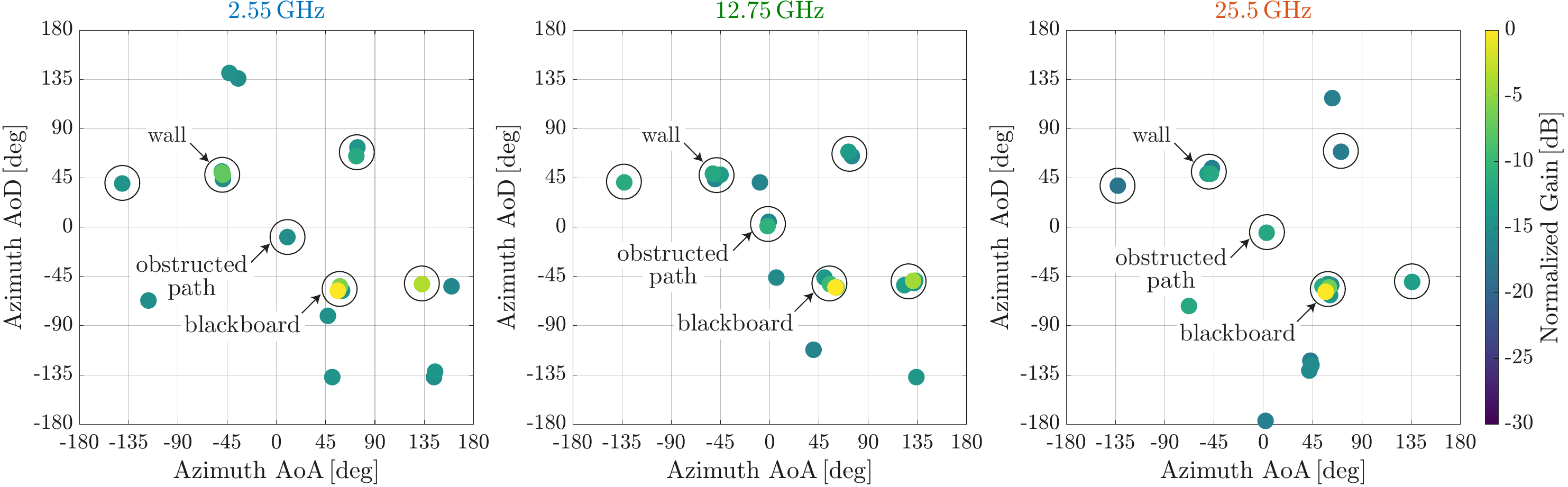}
		\label{fig:SAGE_AoA_vs_AoD_OLOS}
	}

    \caption{Dominant propagation paths exhibit a high degree of angular similarity across the three frequency bands, while weaker MPCs show more pronounced frequency-dependent variations in their relative strength. The common MPCs are highlighted by circles.}
	\label{fig:SAGE_AoA_vs_AoD}
\end{figure*}

Using the measurement setup from~\cref{sec:measurement_setup}, we conduct wireless channel measurements at center frequencies of 2.55\,GHz, 12.75\,GHz and 25.5\,GHz in an indoor lecture room at TU Wien, Vienna, Austria (see~\cref{fig:environment}).
The \ac{Tx} and \ac{Rx} are positioned at the same height of 1.5\,m, representing a \ac{D2D} communication scenario.
The \ac{Tx} remains fixed throughout the campaign, while the \ac{Rx} is placed at 13 different positions according to the grid shown in~\cref{fig:environment}.
Measurements are carried out under both \ac{LOS} and \ac{OLOS} conditions for \ac{Tx}--\ac{Rx} separations of 2, 3, 4, 5 and 6\,m.
At each distance, multiple \ac{Tx}--\ac{Rx} position combinations are considered to capture different propagation conditions and shadowing effects within the environment.
For all three frequency bands, the geometric centers of the virtual transmit and receive \acp{UPA} are kept at exactly the same positions, enabling a consistent comparison of the measured channels across frequency bands (see~\cref{fig:environment}).
In the \ac{LOS} scenario, no physical obstruction is present between the \ac{Tx} and \ac{Rx}, while in the \ac{OLOS} scenario, a whiteboard is placed between the \ac{Tx} and \ac{Rx} to introduce an obstruction (see~\cref{fig:LOS_OLOS_scenarios}).
For all measurements, a bandwidth of 400\,MHz, a frequency spacing of 100\,kHz and a transmit power of 15\,dBm are used.
Since the virtual arrays are synthesized sequentially through mechanical antenna positioning, maintaining a static propagation environment throughout the measurements is essential.
To achieve this, the entire measurement campaign is conducted within a time window during which there is no human presence or movement in the lecture room.
As a result of the measurement campaign, a \ac{MIMO} \ac{CTF} is obtained for each position pair and each of the three frequency bands.

\section{Results} \label{sec:results}
We first post-process the measured channel data. 
A Hann window is applied to the measured \acp{CTF} to suppress spectral sidelobes. 
The windowed \acp{CTF} are then transformed into the delay domain using the \ac{IDFT}, yielding the corresponding \acp{CIR}.
The measured \acp{CIR} contain additive noise, which can bias the estimation of channel characterization parameters. 
To mitigate this effect, we identify and remove the only-noise samples.
The noise floor is estimated using the median-based approach described in~\cite{Sousa1994}. 
A noise thresholding filter is then applied, where the magnitude of each \ac{CIR} sample below the selected threshold is set to zero. 
The threshold is set to 6\,dB above the estimated noise floor for all measurement scenarios, following one of the thresholding approaches evaluated in~\cite{Dupleich2021}.
The \ac{DR} is defined as the difference between the maximum power and the estimated noise floor, increased by the applied threshold. 
Since the resulting \ac{DR} varies across frequency bands and measurement positions, a common dynamic range is used to ensure a consistent comparison across all scenarios. 
Specifically, we limit the \ac{DR} to 30\,dB for all scenarios and set all values outside this range to zero.
Finally, the processed \acp{CIR} are transformed back into the frequency domain using the \ac{DFT}.
Double-directional channel characterization is subsequently performed by extracting the spatio-temporal parameters of the specular \acp{MPC} from the processed \acp{CTF} using a \ac{HRPE} algorithm. 
In this work, we employ the \ac{SAGE} algorithm, following the parameter extraction procedure described in~\cite{Fleury1999}.

The azimuth \ac{AoA} and \ac{AoD} of the 20 extracted specular \acp{MPC}, together with their gains, are shown in~\cref{fig:SAGE_AoA_vs_AoD}.
To keep the analysis concise, only the Rx5 measurement position is considered.
To allow for better comparability between different frequency bands, we normalize the extracted gains such that the strongest \ac{MPC} in each band is set to 0\,dB.
The dominant propagation paths exhibit similar angular positions across all three bands, while weaker \acp{MPC} show greater frequency-dependent variations. 

For the \ac{LOS} case (see~\cref{fig:SAGE_AoA_vs_AoD_LOS}), the strongest component at all three bands corresponds to the direct path, located at $(0^\circ,0^\circ)$.
The reflection from the blackboard is the second strongest component located at approximately $(60^\circ,-60^\circ)$, with its normalized gain increasing from $-6$\,dB at sub-6\,GHz to $-3.5$\,dB at mid-band and $-1$\,dB at \ac{MMW}.
Another shared component originates from the wall opposite the blackboard, located at approximately $(-50^\circ,50^\circ)$, and has a similar gain of about $-13$\,dB across all three bands.

For the \ac{OLOS} case (see~\cref{fig:SAGE_AoA_vs_AoD_OLOS}), the direct path component is attenuated by the whiteboard obstruction, with its normalized gain decreasing to approximately $-15$\,dB at sub-6\,GHz and $-12$\,dB at both mid-band and \ac{MMW}.
Consequently, the strongest resolved component in all three bands corresponds to the reflection from the blackboard at approximately $(60^\circ,-60^\circ)$.
The opposite-wall reflection remains visible, with a normalized gain of approximately $-15$\,dB at sub-6\,GHz and $-12$\,dB at mid-band and \ac{MMW}.

\section{Conclusion} \label{sec:conclusion}
In this paper, we conduct double-directional \ac{MIMO} channel measurements in an indoor environment at 2.55\,GHz, 12.75\,GHz and 25.5\,GHz and analyze the extracted specular \acp{MPC} in the angular domain.
The results demonstrate that the dominant propagation paths exhibit high degree of similarity in angular domain across all three bands, while weaker \acp{MPC} show greater frequency-dependent variations. 
This indicates that the dominant propagation geometry is largely preserved across the considered frequency range, despite the substantial difference in wavelength.

\bibliography{references}
\bibliographystyle{IEEEtran}

\end{document}